\pdfoutput=1
\documentclass{acm_proc_article-sp}
\usepackage{listings}
\usepackage[utf8]{inputenc} 

\usepackage{graphicx}
\usepackage{caption}
\usepackage{subcaption}
\usepackage{courier}

\emergencystretch=\maxdimen
\DeclareCaptionFormat{listing}{\rule{\dimexpr\columnwidth\relax}{0.1pt}\par\vskip1pt#1#2#3}

\begin{document}

\title{SPARQL over GraphX \titlenote{This document is the project report prepared for CS 848. April 2015 }}

\numberofauthors{1} 
%
\author{
%
%
\alignauthor
Besat Kassaie\\
       \affaddr{Cheriton School of Computer Science, University of Waterloo}\\
       \email{bkassaie@uwaterloo.ca}
}

\maketitle
\begin{abstract}
The ability of the RDF data model to link data from heterogeneous domains has led to an explosive growth of RDF data. So, evaluating SPARQL queries over large RDF data has been crucial for the semantic web community. However, due to the graph nature of RDF data, evaluating SPARQL queries in relational databases and common data-parallel systems needs a lot of joins and is inefficient. On the other hand, the enormity of datasets that are graph in nature such as social network data, has led the database community to develop graph-parallel processing systems to support iterative graph computations efficiently. In this work we take advantage of the graph representation of RDF data and exploit GraphX, a new graph processing system based on Spark. We propose a subgraph matching algorithm, compatible with the GraphX programming model to evaluate SPARQL queries. Some experiments are performed to show the system scalability to handle large datasets.

\end{abstract}


\section{Introduction}

Storing web data in the Resource Description Framework (RDF) format \cite{klyne_resource_2004} facilitates linking between heterogeneous domains by not enforcing data to fit into an explicit schema \cite{aluc_workload_2014}. This ability to integrate data from different domains has led to an explosive growth in RDF data recently. RDF, as a standard format for representing Semantic Web datasets, is used in collections such as Linked Open Data \cite{bizer_linked_2009} and DBPedia \cite{Bizer:2009:DCP:1640541.1640848}. In addition, as a data format that can be read and understood by machines, RDF is used for representing data in other domains such as biology \cite{zeng_distributed_2013}. To retrieve data stored in the RDF format, the World Wide Web Consortium (W3C) proposed a standard called SPARQL. Ever growing RDF data mandates providing scalable and efficient systems for answering SPARQL queries. 

There are multiple approaches for storing RDF data and processing SPARQL queries. In some current approaches, RDF data is stored in NoSQL storage engines, relational databases, and native triple stores \cite{ladwig_cumulusrdf:_,harris_4store:_2009}. Answering SPARQL queries, in these systems, is basically performed by relational style joins. However, the graph nature of RDF data incurs a lot of iterative joins, which significantly affects system performance. These iterative joins prevent common data-parallel systems, such as MapReduce to improve the performance considerably. In fact data-parallel systems perform poorly for iterative programming due to excessive disk access. Furthermore, regarding the graph structure of RDF, there are some specialized graph engines which store and process RDF graphs efficiently, such as gStore \cite{Zou:2011:GAS:2002974.2002976} and Trinity.RDF \cite{Zeng:2013:DGE:2488329.2488333}. gStore evaluates SPARQL queries by subgraph matching, while, Trinity.RDF uses graph exploration to answer SPARQL queries.

On the other hand, the enormity of datasets (other than RDF data) which can be represented by graph such as social network data, led the database community to develop scalable and highly efficient graph-parallel processing platforms e.g. Pregel \cite{malewicz_pregel:_2010}, Giraph\cite{_apache_Giraph}, GraphLab\cite{zeng_distributed_2013}, and GraphX\cite{xin_graphx:_2013}. These systems share some properties, such as partitioning graph data for scalability, and also, using a vertex-centric model to implement graph algorithms, which are usually iterative in nature.

Evaluating SPARQL queries can be considered as a subgraph matching problem over an RDF graph. So, potentially we can exploit graph processing systems, specialized for supporting iterative algorithms, to implement the subgraph matching problem using an iterative approach. To the extent of our knowledge, there is a work \cite{Goodman:2014:UVP:2688283.2688287} that uses this potential for evaluating SPARQL Basic Graph Pattern (BGP) queries. However, it has some limitations such as restricting the position of query variables to only subjects or objects .

 The contribution of our work is in evaluating SPARQL queries over GraphX. GraphX is a new graph processing framework which has been developed over Spark, a new dataflow framework which performs up to 100X faster than Hadoop MapReduce. GraphX efficiently combines the advantages of both graph-parallel and data-parallel frameworks. Our approach is based on subgraph matching for answering zero entailment BGP queries.

In the subsequent sections, we present a background on RDF and GraphX. Then, we explain the proposed algorithm and its implementation details. Finally, the evaluation results are presented.

\section{Background}
\subsection{RDF}

The Resource Description Framework (RDF) is a standard format to represent Semantic Web datasets. RDF is commonly used to describe information in such a way that can be understood and read by machines. \textit{Triples} are the core structures of RDF datasets. Each triple consists of a \textit{subject} (s), a \textit{predicate} (p), and an \textit{object} (o). All of these elements , (s, o, p), can be represented by Unique Resource Identifier (URI). In addition, objects can be literals, which are used to represent values such as strings, numbers, and dates. Also, blank nodes are allowed in the position of \textit{s} and \textit{p}. In this work, blank nodes are supported in BGP queries and are treated similarly to variables.

\begin{figure}[h!]
 
  \centering
    \includegraphics[width=0.4\textwidth]{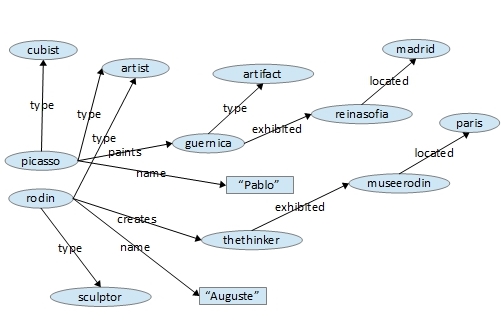}
     \caption{RDF graph (adopted from \cite{kaoudi_rdf_2014})}
     	\label{fig:RDFGraph}
\end{figure}

\begin{figure}[h!]
 
  \centering
    \includegraphics[width=0.25\textwidth]{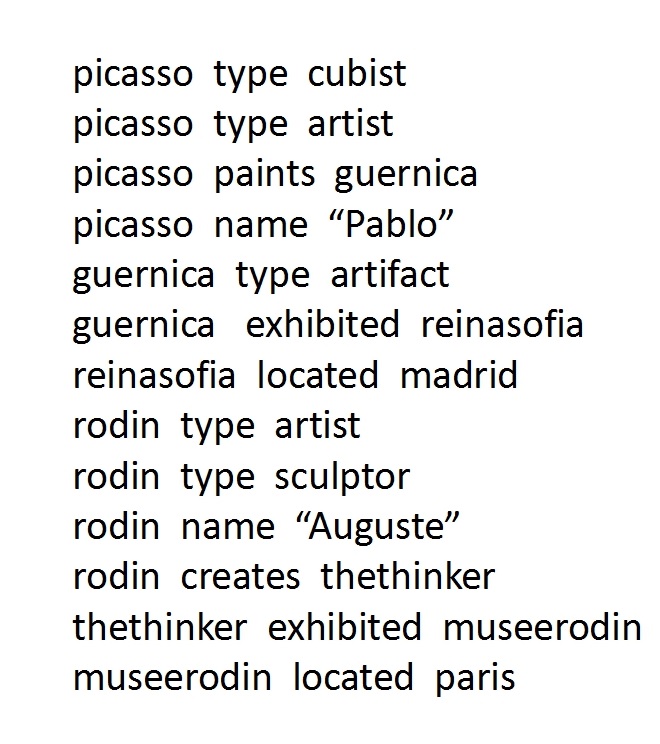}
     \caption{N-Triples syntax (adopted from \cite{kaoudi_rdf_2014})}
     	\label{fig:NTripple}
\end{figure}

By using URI to refer to concepts, relationships, or entities, RDF covers the linking structure of the Web. The RDF structure forms a directed, labeled graph where the subjects and the objects are considered as vertices of the graph. Also, for labelling the directed edges between subjects and objects, the predicates are used. Modelling RDF data as graphs, allows to use graph processing systems to efficiently extract information from RDF data. Figure~\ref{fig:RDFGraph} depicts a sample RDF graph. There are various syntaxes for creating and exchanging RDF data such as RDF/XML, N-Triples, and Turtle. Figure~\ref{fig:NTripple} shows the RDF graph of Figure~\ref{fig:RDFGraph} in the N-Triple representation.

\begin{figure*}
	\centering
	\begin{subfigure}[b]{0.18\textwidth}
		\includegraphics[width=\textwidth]{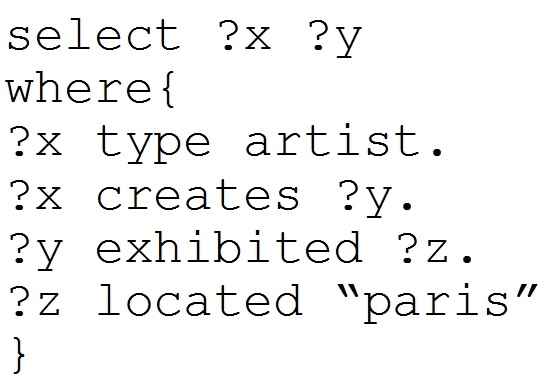}
		\caption{}
		\label{fig:Query}
	\end{subfigure}%
	\qquad
	~ 
	\begin{subfigure}[b]{0.45\textwidth}
		\includegraphics[width=\textwidth]{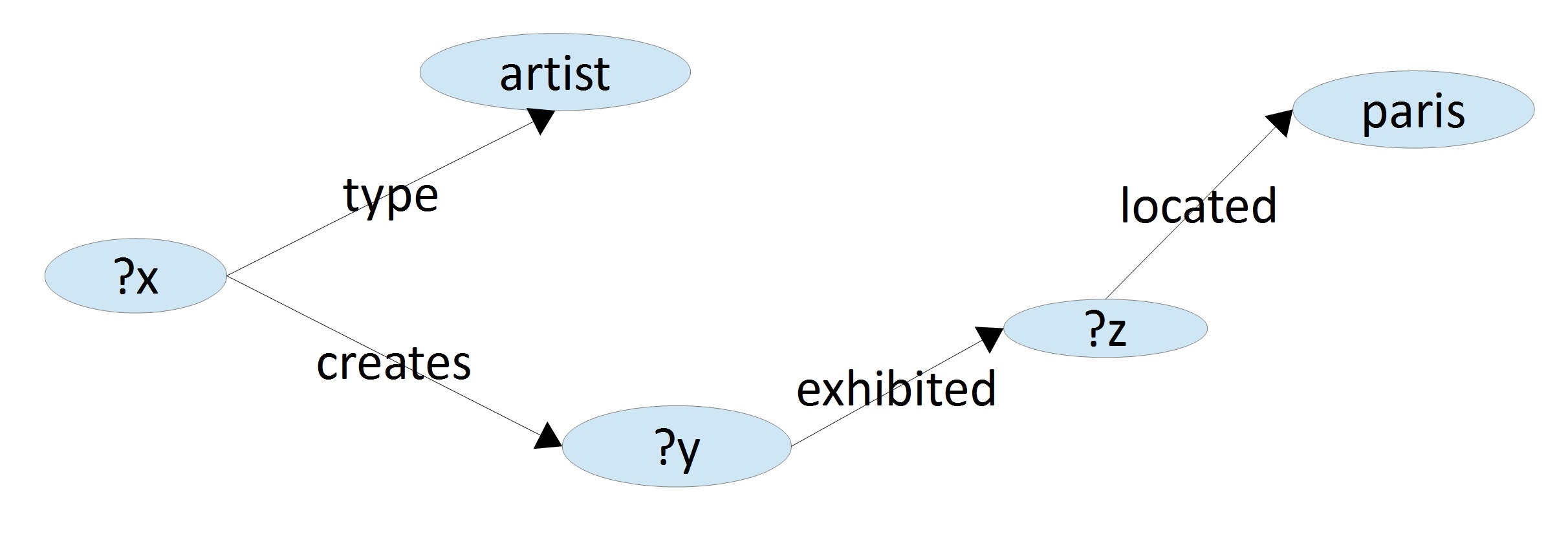}
		\caption{ }
		\label{fig:QueryGraph}
	\end{subfigure}
	\caption{A simple SPARQL query (a) and its query graph (b)}\label{fig:Query_Graph}
\end{figure*}

SPARQL is a SQL-like query language in which variables are indicated by ‘\textit{?}’ prefix. SPARQL covers a wide range of operators over graphs such as filters, negations, and unions, among others. Our aim in this paper is to cover a fundamental subset of SPARQL queries called Basic Graph Pattern (BGP). A sample of BGP and its query graph are shown in Figure~\ref{fig:Query_Graph}. The building blocks of SPARQL queries are \textit{triple patterns}, e.g. \textit{?x type artist}. One or more triple patterns represent a query graph.

There are two basic types of BGP queries. The first type is \textit{star-join} in which a number of triple patterns in a query share the same variable in the subject position. Furthermore, when the object variable of a triple plays a role as the subject variable of the next triple (it might happen for multiple consecutive triples) a \textit{chain} query is created. The sample query, shown in Figure~\ref{fig:Query_Graph}, is composed of one star-join and one chain query.

\subsection{GraphX}

GraphX is a lightweight graph processing library on top of Spark, a general-purpose data flow framework. It achieves the benefits of specialized graph processing frameworks in a general-purpose distributed dataflow system by adopting the bulk-synchronous model of computation. Furthermore, GraphX proposes a single system which supports the entire pipeline of graph analytics, including data preprocessing, graph processing, and post data analysis.

Apache Spark, as a dataflow system, has some important features which make this system suitable for implementing GraphX. Resilient Distributed Dataset (RDD) is the storage abstraction in Spark which can keep data in memory. This increases the efficiency of iterative algorithms over graphs. Furthermore, Spark users can define their own partitioning algorithms over RDDs, which is essential for encoding partitioned graphs. Also, Spark is a fault tolerant system. Spark achieves that by keeping the lineage of operations executed for building an RDD. This lineage will be used for reconstructing RDDs in case of any failure. The lineage graph is very small and does not affect the system performance. In addition, Spark provides users with a high-level and easily extendable API.

GraphX represents a graph internally as a pair of RDD collections: a vertex collection which stores vertex properties along with a unique key (\textit{VertexRDD}) and an edge collection which contains properties of edges along with their corresponding identifiers of the source and destination vertices (\textit{EdgeRDD}). By this representation, GraphX supports many of common computations on graphs which other graph processing systems do not easily support, such as adding new properties to vertices, as well as, analyzing the results of graph computations by some collection operators such as \textit{join operator}.

GraphX allows graphs with common vertex and edge collections to use the same index data structure. Sharing the index data structure improves performance and decreases storage consumption. By representing graphs as collections of vertices and edges, graph-parallel computations can be implemented by a series of \textit{join} stages and \textit{group-by} stages along with \textit{map} operations.

\begin{figure}[h!]
 
  \centering
    \includegraphics[width=0.2\textwidth]{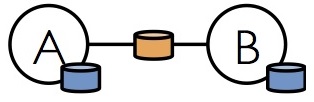}
     \caption{Triplet View}
     	\label{fig:tripletview}
\end{figure}

The triplet view (Figure~\ref{fig:tripletview}) is the building block of graph-computation in GraphX. A triplet view is composed of an edge and its related source and destination vertex properties. To create the triplet view, in a join stage, GraphX performs a three-way join on vertex and edge properties. Using triplet views, the neighborhood of vertices are constructed easily. Furthermore, aggregations are computed by using a group-by stage on either source or destination of the triplet views. In fact, a graph processing model such as GAS (Gather, Apply, Scatter) can also be expressed by these two stages, i.e., the group-by stage with some map operations, and the join stage are equivalent to gathering messages, apply function, and scattering messages, respectively.

GraphX offers some specialized graph operators, among them, Listing ~\ref{lst:operators} (Appendix.\ref{App:AppendixA}) shows some of GraphX operators. By vertex, edge, and triplet operators, properties of vertices and edges, and also, the triplets view can be retrieved. The \textit{aggregateMessages} operator basically applies the map and group-by operators on the triplets view. Users define two functions to be executed on each triplet view: \textit{sendMsg} and \textit{mergeMsg} functions. The sendMsg function is basically a map function which is applied to each triplet and sends some messages to either source or destination vertices of the triplet. The mergeMsg is a binary reduce function, which gathers the message values, received from sendMsg functions, at the source and destination vertices. These two user-defined functions are used by the aggregateMessages operator to produce a new VertexRDD containing the required results. Another set of useful operators are join operators which can be used for updating properties of edges or vertices. An iterative computation over a graph can be implemented by a sequence of using aggregateMessages to produce new values for vertices and then applying join to include the new values into properties of vertices in a graph.

From a systemic point of view, GraphX partitions the vertex collection based on the hashed value of vertex ids. To achieve efficient joins over the vertices, a local hash index is stored in each partition. To have a highly reusable index, there is a bitmask within each partition determining whether a vertex is deleted. Edges can be partitioned by a user-provided function. GraphX also supports horizontal vertex-cut partitioning over the edge collection to reduce the communication overhead in natural graphs.

Being RDDs, vertex and edge collections are immutable objects in GraphX. So, graph operators create a new RDD whenever an existing RDD is modified. The RDD immutability reduces the memory overhead by increasing the reusability of indices. Furthermore, by immutability, GraphX achieves faster joins when it combines the VertexRDD returning from an aggregateMessages operator with the original graph's VertexRDD. This is beneficial in iterative operations because the new and old VertexRDDs share the same index. GraphX reuses indices whenever it is possible, i.e., if an operator does not change or restricts the graph structure, the index can be reused.

GraphX proposes efficient mechanisms for creating and maintaining the triplets view. To construct the triplets view, GraphX performs a three-way join on the source and destination vertex properties and the edge properties. Since the edges and vertices are partitioned separately, data movement is necessary. Graphx sets the edge partitions as \textit{join sites} and moves the vertices over the network. Usually, there are a small number of vertices in comparison with the number of edges, and also, many edges of a single vertex possibly are located in the same partition. Because of these two reasons the vertex movement process, called \textit{vertex mirroring}, reduces the communication overhead dramatically.

In addition, GraphX keeps the information about the edge partitions of the adjacent edges for each vertex. This information is co-partitioned with each vertex and is used for an efficient multicast join. In the multicast join, vertex properties are sent only to corresponding edge partitions containing adjacent edges. As another optimization, GraphX supports the Incremental View Maintenance mechanism in which only the modified vertices are re-routed to the join sites and the unchanged local vertices are reused. This mechanism efficiently prevents reconstructing the entire triplet view from scratch after each iteration.

\section{SOLUTION OVERVIEW}
\begin{figure}[h!]
	
	\centering
	\includegraphics[width=0.45\textwidth]{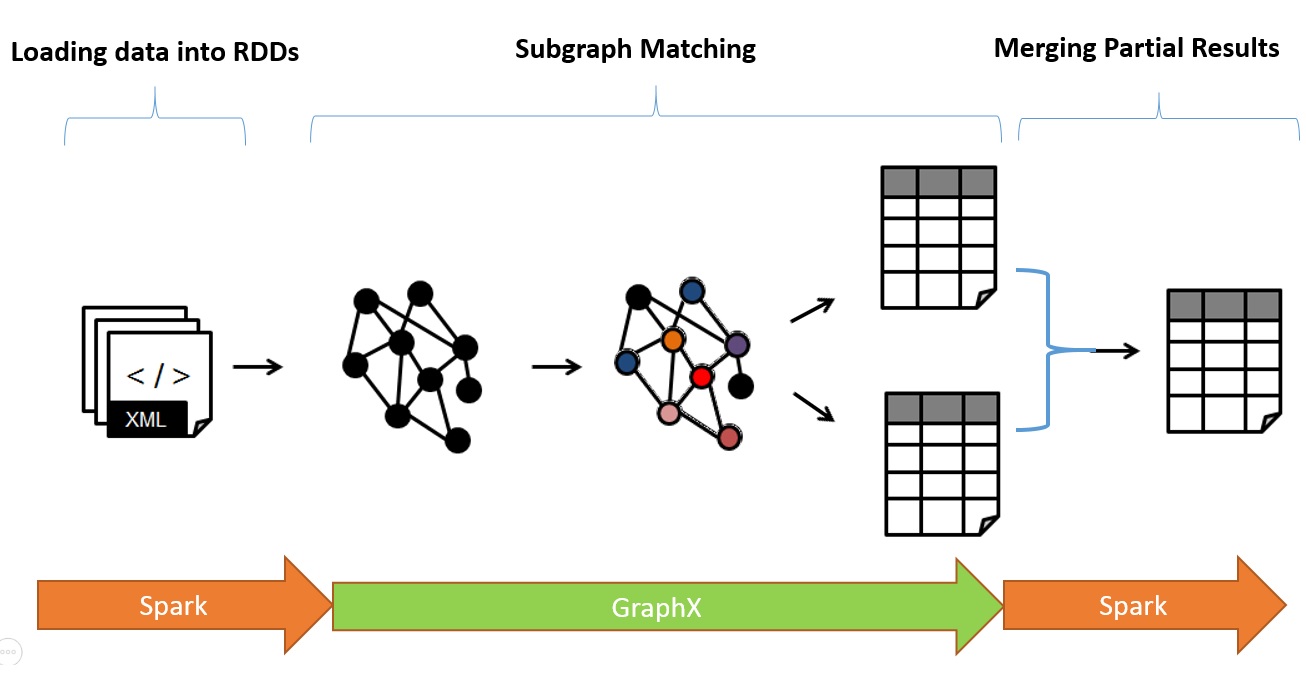}
	\caption{Solution Pipeline}
	\label{fig:pipeline}
\end{figure}

Our solution relies on some important properties of GraphX. GraphX is implemented on top of the Spark framework and stores graph data as Spark in-memory RDDs. This storage abstraction is well suited for our algorithm, which is iterative in nature. The second property is that GraphX views a graph as a collection of triplets. This view is compatible with the RDF data model, which represents data as triples of subject, object, and predicate. Furthermore, unlike other graph processing systems, GraphX does not restrict data representation to the graph view. GraphX provides both the collection view and the graph view of the same physical data. Each view has its own operators relevant to the semantics of the view. Moreover, in GraphX, switching between these views does not impose overhead in terms of data movement and duplication. In this work, we need the graph view for the iterative pattern matching and the collection view for merging partial results of subgraph matching. 
Figure \ref{fig:pipeline} shows our overall solution pipeline and switchings between the collection view (Spark) and the graph view (GraphX). 

\subsection{Subgraph Matching Algorithm}

\begin{figure*}
	\centering
	\begin{subfigure}[b]{0.2\textwidth}
		\includegraphics[width=\textwidth]{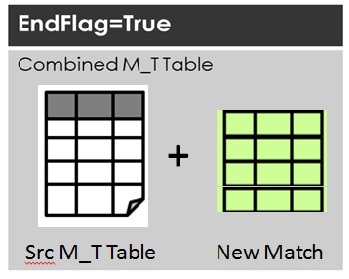}
		\caption{}
		\label{fig:dest}
	\end{subfigure}%
	\qquad
	~ 
	\begin{subfigure}[b]{0.19\textwidth}
		\includegraphics[width=\textwidth]{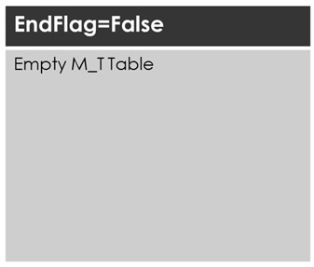}
		\caption{ }
		\label{fig:src}
	\end{subfigure}
	\caption{Message structures of the destination(a) and source (b) vertices} \label{fig:message}
\end{figure*}

Subgraph matching is the central part of our solution which works on a graph view of RDF data. For this graph view, we need to specify the properties of vertices and edges. We assign three properties to each vertex, i.e., a \textit{label}, a \textit{Match\_Track} table (M\_T), and an \textit{endFlag}. The label keeps the value of its corresponding subject/object. The M\_T table contains variable and constant mappings in paths which end in each vertex. The endFlag shows whether a vertex is located at the end of a path. Paths are made of a sequence of matched BGP triples. We call a vertex whose endFlag is set to \textit{true}, an \textit{end vertex}. Also, each edge has a property called \textit{edge label}. The value of an edge label is its corresponding predicate value.

\begin{lstlisting}[caption={Subgraph Matching Algorithm},label={lst:overall_alg}]
def doMatch()
{
 while (!BGPinOrder.isEmpty)
 {val CurrentBGPTriple = BGPinOrder.dequeue
  val vertexSets = FindCandidateVertices()
  val newVertexVals = mygraph.aggregateMessages(sendMsg,mergeMsg)
  //Merge the newVertexVals with the old properties
  mygraph = mygraph.joinVertices(newVertexVals,joinMapper)}
 Final_RDD_MTs = GetMTFromEndVertices()
 FinalResult = Join(Final_RDD_MTs)
 }
\end{lstlisting}

Listing ~\ref{lst:overall_alg} represents the main function of our subgraph matching algorithm. This function iterates over all BGP triples in a SPARQL query. So, the number of iterations in our algorithm is determined by the number of BGP triples in the query. We assume that each BGP query represents a connected subgraph. Based on this assumption, \textit{BGPinOrder}, a list of BGP triples, has to be ordered in such a way that the sequence of triples preserves the connectivity of the BGP triples, i.e., each BGP triple has a common variable with at least one of the previous triples in the BGPinOrder list. Furthermore, our algorithm performance is sensitive to the order in which the triples are evaluated. Although there might be several orders that preserve connectivity in BGPinOrder, only some of them are optimal. However, currently, we do not focus on finding the optimal orders.

\begin{lstlisting}[caption={sendMsg function},label={lst:sendMSG_alg}]
def sendMsg(triplet){
if(SrcCandidateSet.contains(triplet.src) && DestCandidateSet.contains(triplet.dest))
 {  val mappedConstants = matchConstantParts(triplet,currentBGPTriple)
   if(allConstantsMapped)
    {val mappedVariables = matchVariableParts(triplet,currentBGPTriple)    
     if(allVariablesMapped)
      {//mapList is the combination of mappedVariables and MappedConstants
       var mapList = mappedVariables++mappedConstants    
       if(!triplet.src.M_T.isEmpty)
       {//add mapList to each of rows in M_T Table of src
       mapList= addtoCurrentM_T(triplet.src.M_T,mapList)}
       if (!triplet.dest.M_T.isEmpty)
       {
       //joining mapList with M_T table of dest
       	mapList=mapList.Join(triplet.dest.M_T)
       }
       //the endFlag of dest will be set to true   
       var DestMsg=(mapList,true)
       //the endFlag of src will be set to true   
       var SrcMsg=(emptyMapList,false)
      //messages to src and destination
      triplet.sendToDst(DestMsg)
      triplet.sendToSrc(SrcMsg)}
     }
   else
   {val SrcMsg=(emptyMapList,false)
    triplet.sendToSrc(SrcMsg) }
    }
 }
\end{lstlisting}

In our work, we take advantage of the triplet view in GraphX by using the \textit{aggregateMessages} operator. This operator first applies a user-defined \textit{sendMsg} function, in parallel, to each graph triplet. In this sense, the sendMsg function could be considered as a map function that matches the \textit{CurrentBGPTriple} with all graph triplets. If a match is found, the sendMsg function prepares and sends different messages to the destination and the source vertex of the triplet. Then, using the mergeMsg function, as a reduce function, the received messages are aggregated at their destination vertex. The outcome of the aggregateMessages operator is a new VertexRDD which contains new vertex properties.

As the last step in each iteration, we use the GraphX \textit{joinVertices} function to combine the old property values with the new values in each vertex. As RDDs are immutable entities, this operation creates a new graph in which vertex property values are updated. The \textit{joinMapper} is a user-defined function by which we can control how the old and new property values will be combined. The entire process is repeated for the next BGP triple in the SPARQL query.

\begin{lstlisting}[caption={mergeMsg function},label={lst:mergeMsg_alg}]
def mergeMsg(a, b)
{ //append M_T tables vertically in a vertex
  val M_T=a.M_T ++ b.M_T
 //determines the endFlag
  val endFlag	=a.endFlag || b.endFlag
  return (M_T,endFlag)}
\end{lstlisting}

In this triple matching process, it is possible to generate some partial subgraphs which are not valid, i.e, do not belong to the final subgraphs. To reduce memory usage and to prevent unnecessary computations, in each iteration, the \textit{FindCandidateVertices} function determines the candidate vertices for each variable of CurrentBGPTriple, based on M\_T tables of the end vertices. M\_T tables keep the previous variable-vertex mappings. For a variable which has been mapped to some vertices previously, the candidates will be limited to those vertices. If a variable has not been mapped to any vertices so far, all graph vertices are candidates. 

After evaluating all BGP triples, we join the final M\_T tables of the end vertices, which contain partial results, to generate query answers. The join operator on RDDs is supported internally in Spark. So, we take advantage of the GraphX ability to switch to a collection view of vertices to extract their M\_T tables as a collection of RDDs. These RDDs are joined based on their common variables. We present more implementation details in the subsequent sections.

\subsubsection{sendMsg}

GraphX executes sendMsg, a user-defined function, on all graph triplets in parallel (Listing ~\ref{lst:sendMSG_alg}). The sendMsg function matches the constants of the current BGP triple with a given triplet. If all constants are matched, then mapping for triple variables is determined. This is a valid vertex-variable mapping if the mapped vertices belong to the candidate vertices of the variables, determined by the FindCandidateVertices function.

If a mapping exists, the next step is to send appropriate messages to the source and destination vertices of the triplet. The structure of the messages for the source and destination vertices are shown in Figures~\ref{fig:message} (a-b). The message to the destination vertex contains a combination of the new vertex-variable mapping and the M\_T table of the source vertex. This combination is performed by horizontally appending the new mapping to each row of the M\_T table. This keeps the track of current matchings in the destination vertex. Next, we need to perform a join operation between the new mapping and the destination M\_T table, in case that M\_T table is not empty. This handles connecting two paths in queries with cyclic patterns. We also need to inform the destination vertex to set its endFlag to true. The structure of the message to the source vertex contains a false value for endFlag and an empty M\_T table to remain the M\_T table of the source vertex unchanged. 

\begin{figure*}
	\centering
	\begin{subfigure}[b]{0.40\textwidth}
		\includegraphics[width=\textwidth]{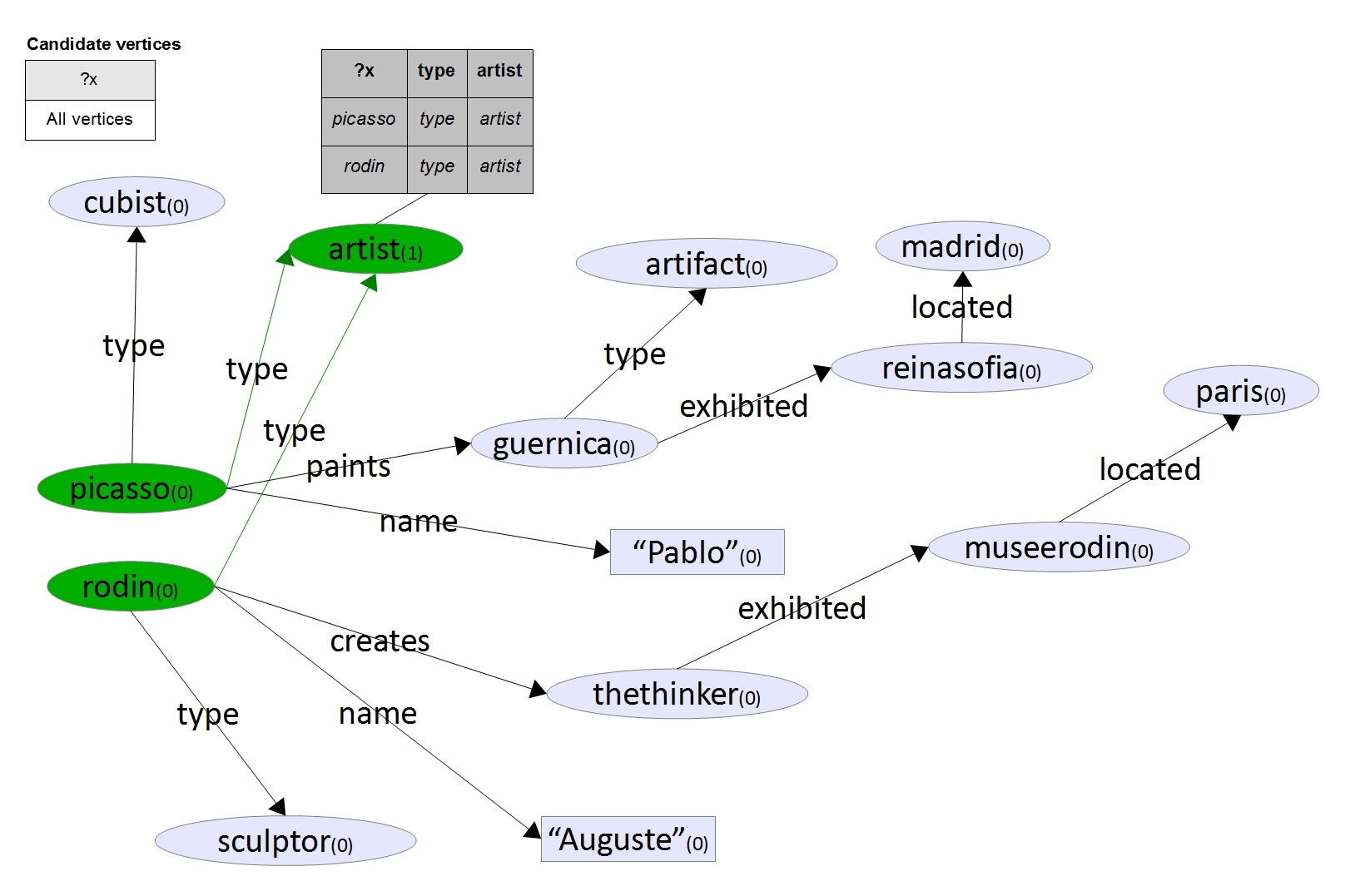}
		\caption{Iteration 1: evaluating \textit{?x type artist}}
		\label{fig:Iterration1}
	\end{subfigure}%
	\qquad
	~ 
	\begin{subfigure}[b]{0.40\textwidth}
		\includegraphics[width=\textwidth]{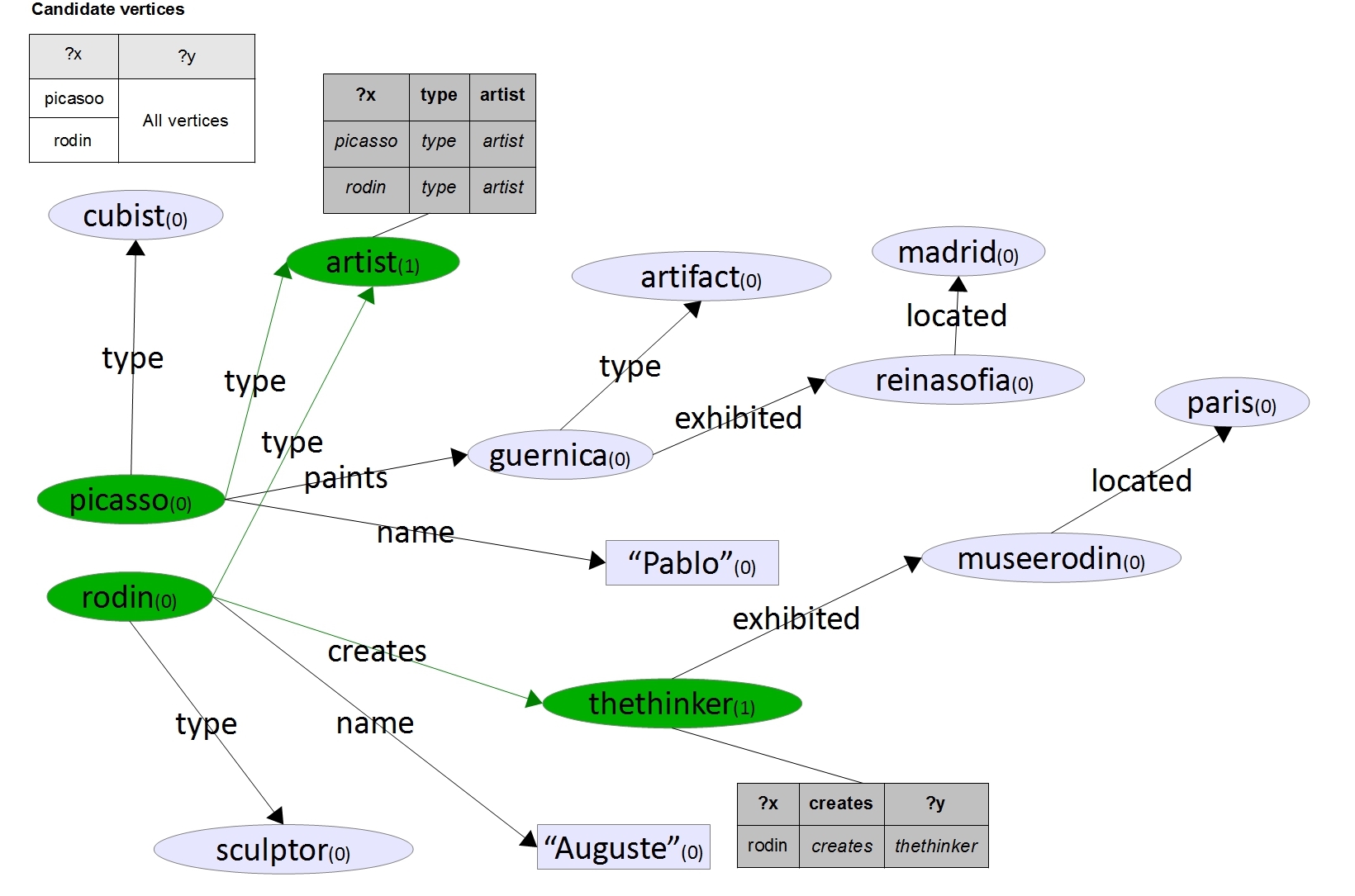}
		\caption{Iteration 2: evaluating \textit{?x creates ?y}}
		\label{fig:Iterration2}
	\end{subfigure}
	
	~ 
	\begin{subfigure}[b]{0.40\textwidth}
		\includegraphics[width=\textwidth]{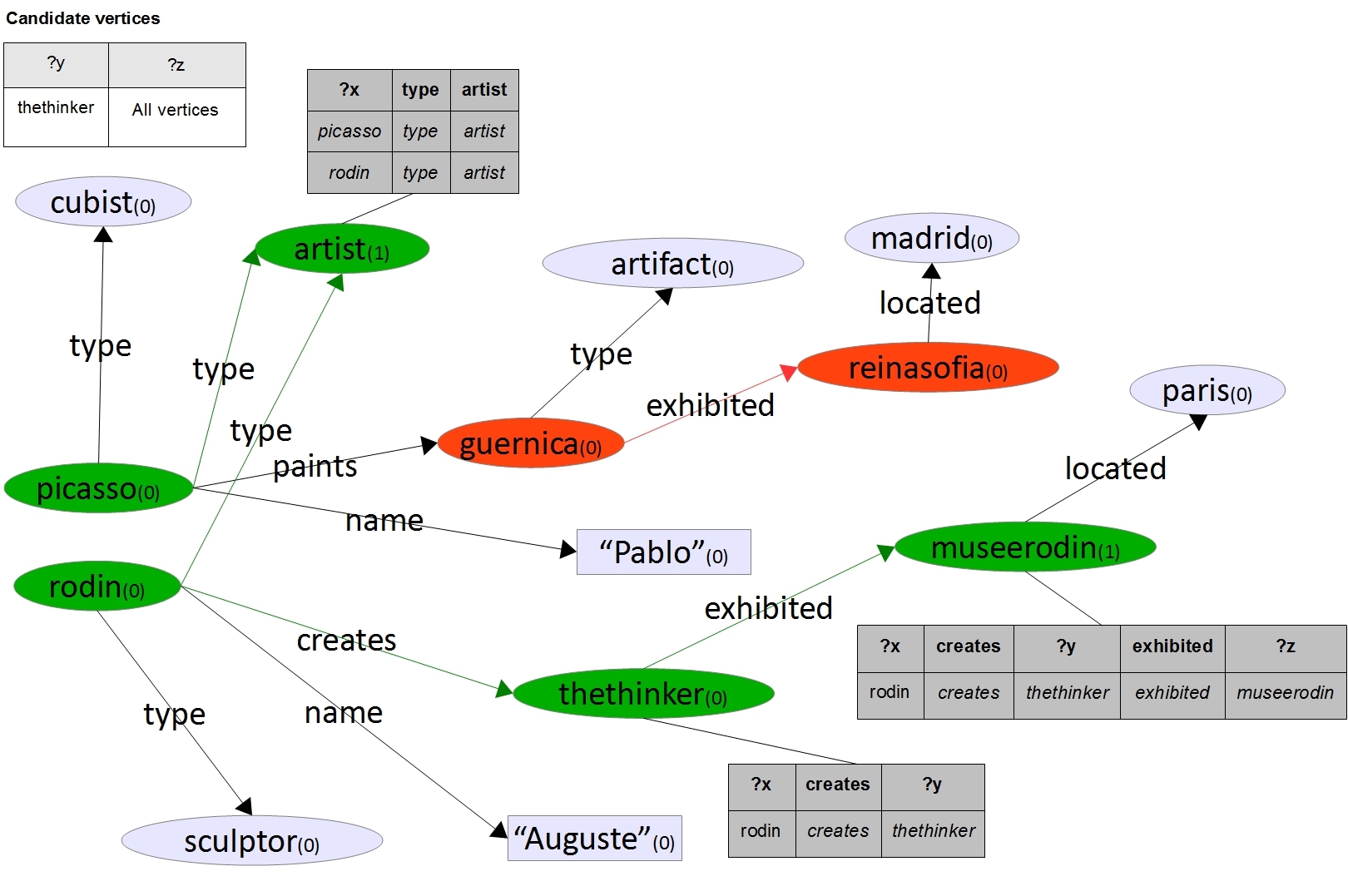}
		\caption{Iteration 3: evaluating \textit{?y exhibited ?z}  }
		\label{fig:Iterration3}
	\end{subfigure}
	\qquad
		\begin{subfigure}[b]{0.40\textwidth}
			\includegraphics[width=\textwidth]{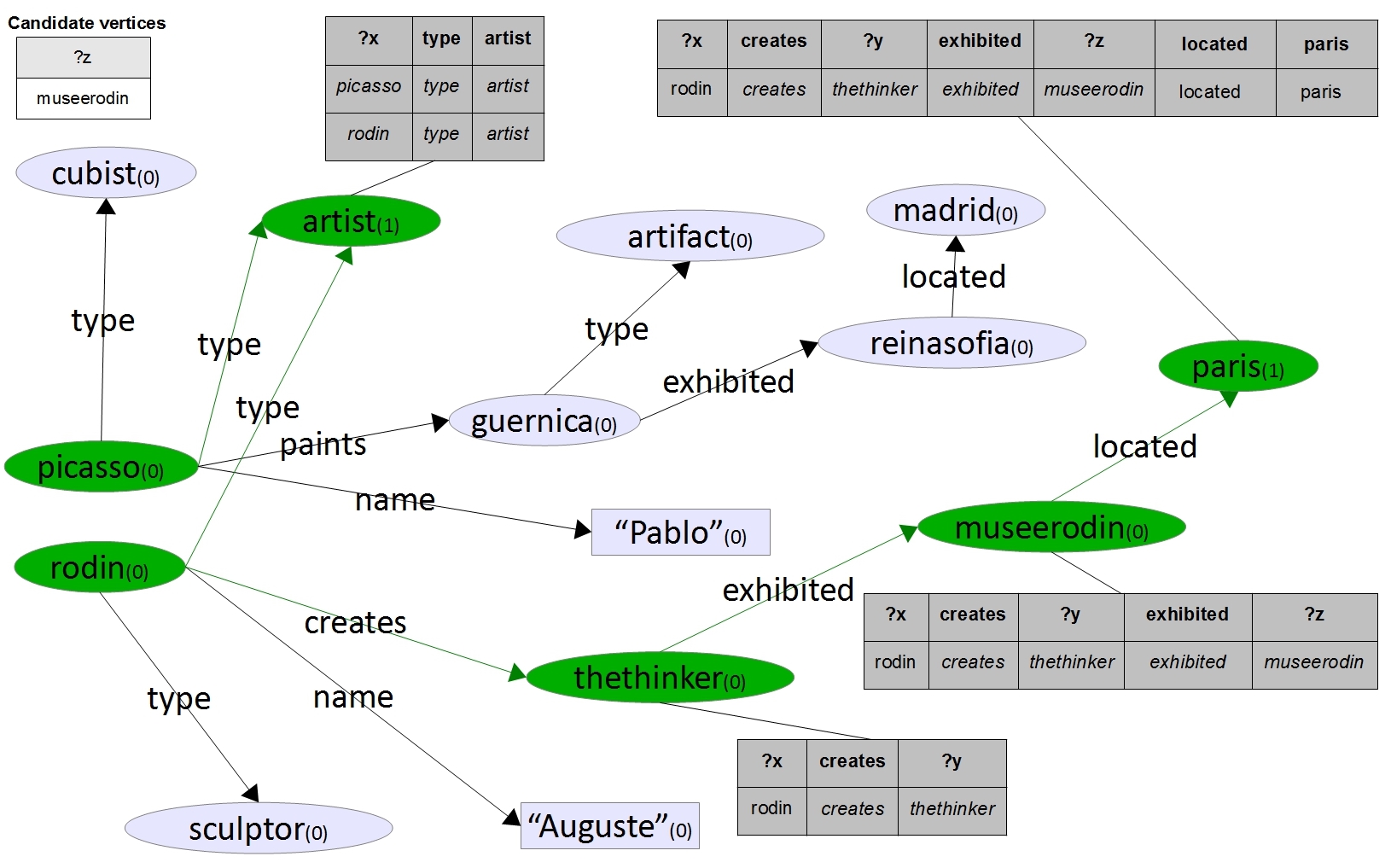}
			\caption{Iteration 4: evaluating \textit{?z located "paris"}}
			\label{fig:Iterration4}
		\end{subfigure}
	\caption{Running the proposed subgraph matching algorithm on the sample graph. The dark gray tables and values inside parentheses represent M\_T tables and endFlags respectively for each iteration}\label{fig:example}
\end{figure*}

\begin{lstlisting}[caption={Join mapper function},label={lst:joinMapper_alg}]
def joinMapper(vertexId,oldValue,newValue)
{ //replaces old M_T if new M_T is not empty
  var newM_T=oldValue.M_T
  if (newValue.M_T.size>0)
  newM_T=newValue.M_T
 //returns the new vertex with updated properties
  return(vertexId,newM_T,newValue.endFlag)}

\end{lstlisting}

If a mapping fails while the source is a candidate vertex, a message is sent to the source vertex with the earlier structure to set its endFlag to \textit{false}. This prevents building up new chains over that vertex. Otherwise, no message will be sent to the source and destination vertex.

\subsubsection{mergeMsg}

GraphX executes mergeMsg (Listing ~\ref{lst:mergeMsg_alg}), a user-defined function, on all vertices in parallel and combines the messages sent to a vertex. The message combination works by vertically appending M\_T tables of all the messages and performing OR operation over the received endFlags. So, the aggregated message sent to the vertices which are mapped to an object, contains the accumulated track of the match sequences that ends in those vertices. As well, this message informs these vertices to set their endFlag to \textit{true}. For the vertices mapped to a subject, the aggregated message has an empty M\_T table. Also, it informs those vertices, that they are not end vertices any more by setting their endFlag to \textit{false}.

\subsubsection{JoinMapper}
The aggregateMessages operator, returns a VertexRDD that contains new properties for each vertex. So, the next step is to update current properties of graph vertices with these new properties. The JoinMapper function (Listing ~\ref{lst:joinMapper_alg}) specifies how to combine the new and old properties. The properties consist of M\_T tables and endFlags. The new M\_T table replaces the old M\_T table if new M\_T table is not empty. Also, the old value of endFlag is replaced with the new one.

\subsubsection{Running Example}

\begin{figure}[h!]
 
  \centering
    \includegraphics[width=0.4\textwidth]{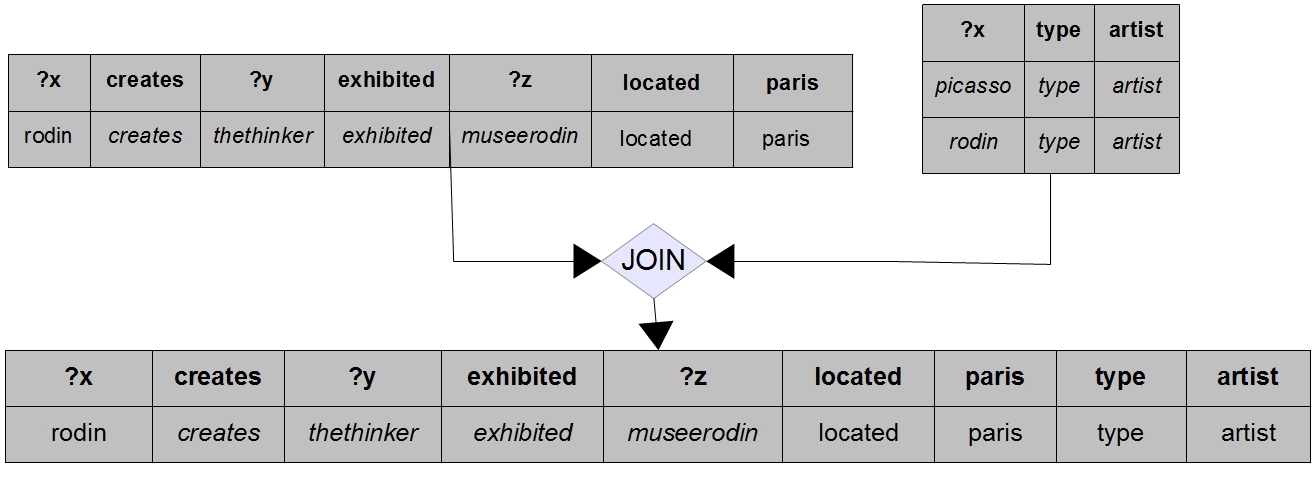}
     \caption{Joining the M\_T tables of end vertices: "paris" and "artist"}
     	\label{fig:Running_Example_join}
\end{figure}

In this section we run the proposed algorithm to answer the simple query shown in Figure~\ref{fig:Query_Graph} over the RDF graph of Figure ~\ref{fig:RDFGraph}. There are four iterations for this query graph because it contains four BGP triples. Figures ~\ref{fig:example}(a-d) show the entire process and the graph state including M\_T tables, endFlags, and the candidate vertices list, after each iteration. In Figure~\ref{fig:example}(c), the red triple shows a possible match, however it is not a valid mapping since the candidate vertices list of \textit{?y} does not contain the \textit{“guernica”} vertex. Figure~\ref{fig:Running_Example_join} represents how the final results are generated by the Spark join operation over the M\_T tables of two end vertices: \textit{“paris”} and \textit{“artist”}. 

\section{Evaluation}

In this section we present the experiments conducted to evaluate our system. We use the \textit{Lehigh University Benchmark} (LUBM) \cite{LUBM_2005} datasets and queries. This benchmark generates synthetic data with arbitrary sizes. In this work, we use four different data sizes; LUBM1 (103,397 triples), LUBM5 (646,128 triples), LUBM10 (1,316,993 triples), and LUBM20 (2,782,419 triples). We use seven representative queries (Appendix \ref{App:AppendixB}), which also are used in \cite{zou_gstore:_2013, Yuan_2013,Atre_2010}. 

We run the experiments on a cluster of 6 machines. Each machine has \textit{AMD Opteron(tm) Processor 4174 HE 2.3 GHz} with 12 cores and 14 GB of RAM. To manage the cluster, we use Spark in the \textit{standalone} mode. For each experiment, one of the cluster machines is the designated master and the other machines play worker role. We run each experiment three times and present the average values in the plots.

\begin{figure*}
	\centering
	\begin{subfigure}[b]{0.45\textwidth}
		\includegraphics[width=\textwidth]{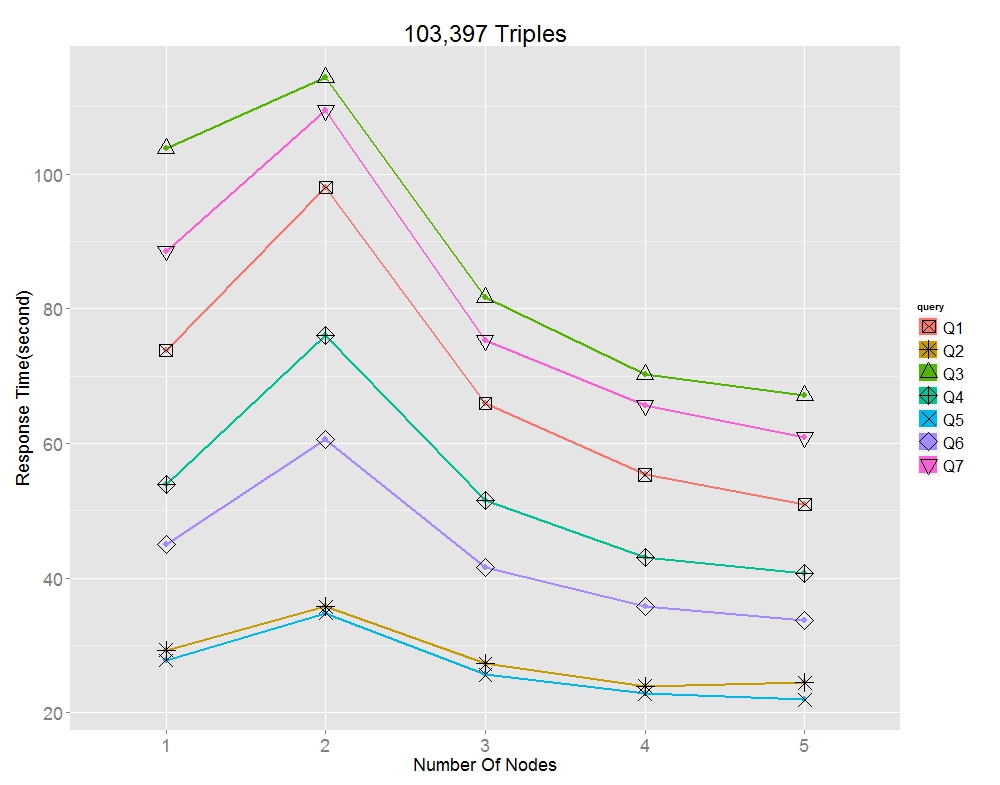}
		\caption{}
		\label{fig:LUBM1}
	\end{subfigure}%
	\qquad
	~ 
	\begin{subfigure}[b]{0.45\textwidth}
		\includegraphics[width=\textwidth]{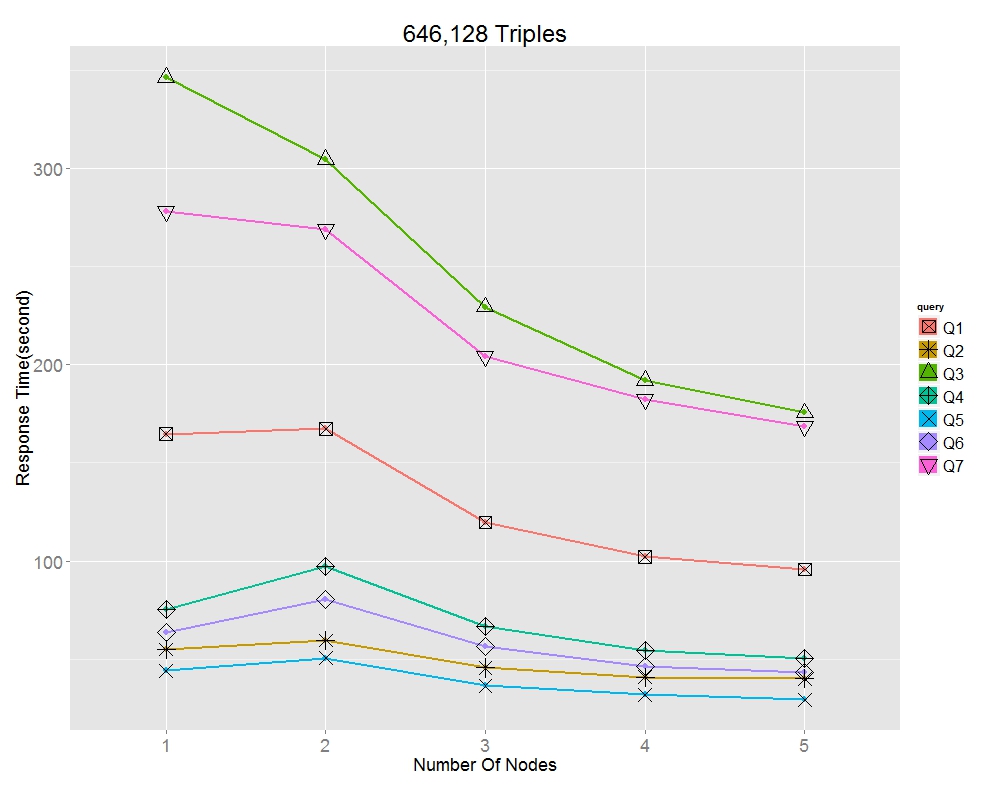}
		\caption{}
		\label{fig:LUBM5}
	\end{subfigure}
	
	~ 
	\begin{subfigure}[b]{0.45\textwidth}
		\includegraphics[width=\textwidth]{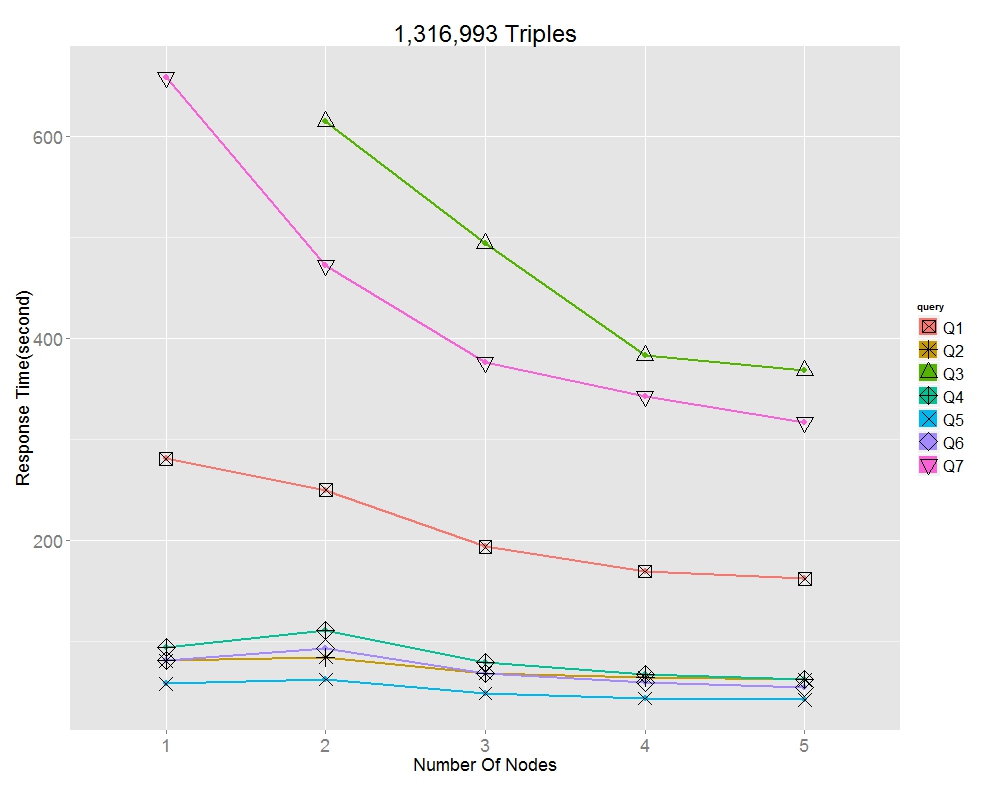}
		\caption{}
		\label{fig:LUBM10}
	\end{subfigure}
	\qquad
		\begin{subfigure}[b]{0.45\textwidth}
			\includegraphics[width=\textwidth]{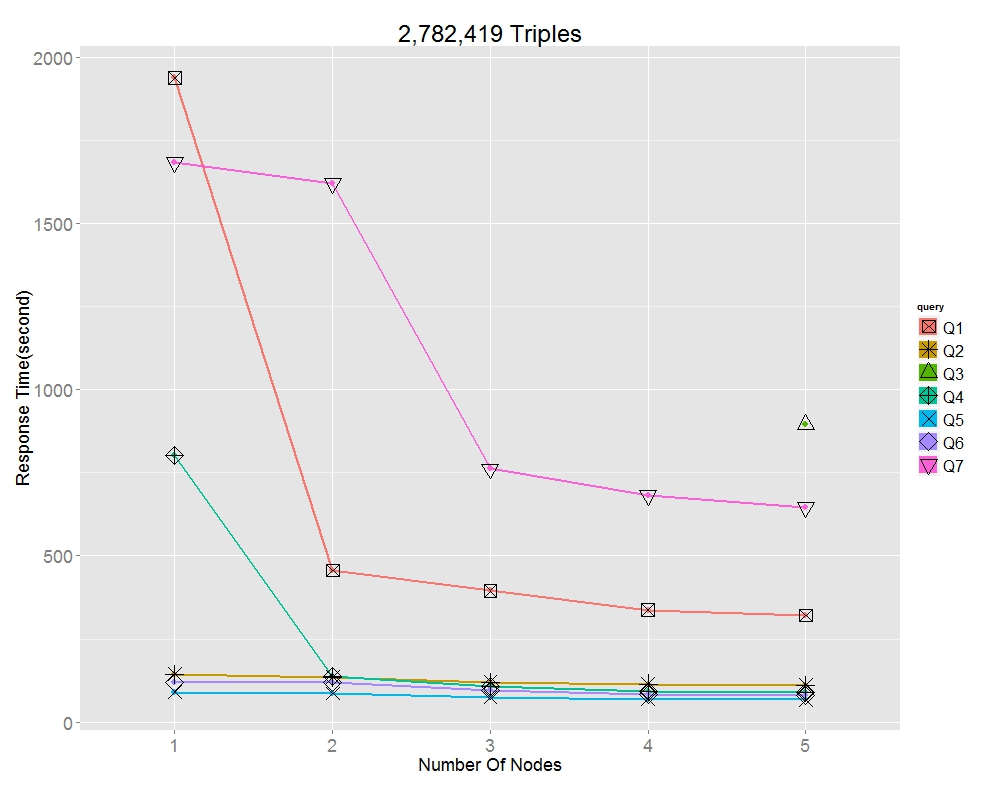}
			\caption{}
			\label{fig:LUBM20}
		\end{subfigure}
	\caption{Query response time for different number of cluster nodes}\label{fig:scalability}
\end{figure*}

Figures \ref{fig:scalability} (a-d) show the scalability of the proposed method over different data sizes. In each plot the effect of varying the number of cluster nodes on the system response time is presented. Generally, the query response time decreases as the number of nodes increases. However, this reduction is more significant for larger datasets and complex chain queries (Q1, Q3, and Q7). 

For small data sizes, by increasing the number of nodes the response time decreases less significantly. In fact, the response time increases for LUBM1 when we increase the number of nodes from one to two (Figure \ref{fig:LUBM1}). The reason is that the effect of having two nodes on reducing computation cost is less significant than the increased communication cost due to RDD shuffling. RDDs are shuffled when new properties of vertices are combined with their old values. This is also the reason of less significant reduction in response time for other small data sizes.

\begin{figure*}
	\centering
	\begin{subfigure}[b]{0.45\textwidth}
		\includegraphics[width=\textwidth]{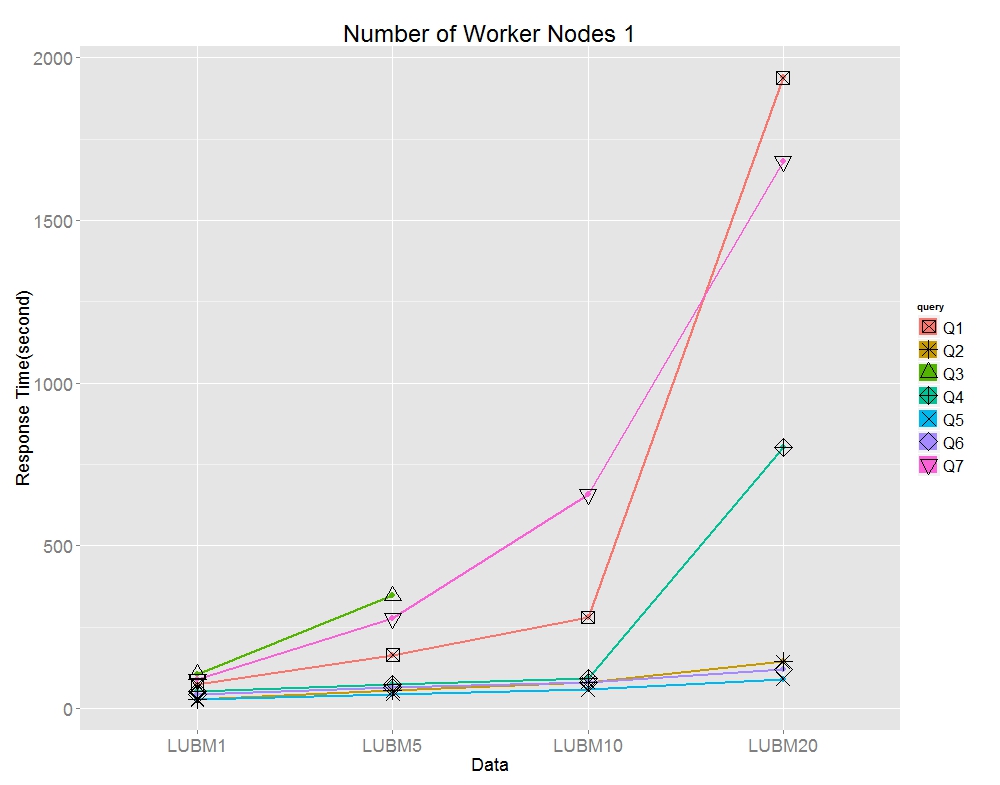}
		\caption{}
		\label{fig:Node1}
	\end{subfigure}%
\qquad
	~ 
	\begin{subfigure}[b]{0.45\textwidth}
		\includegraphics[width=\textwidth]{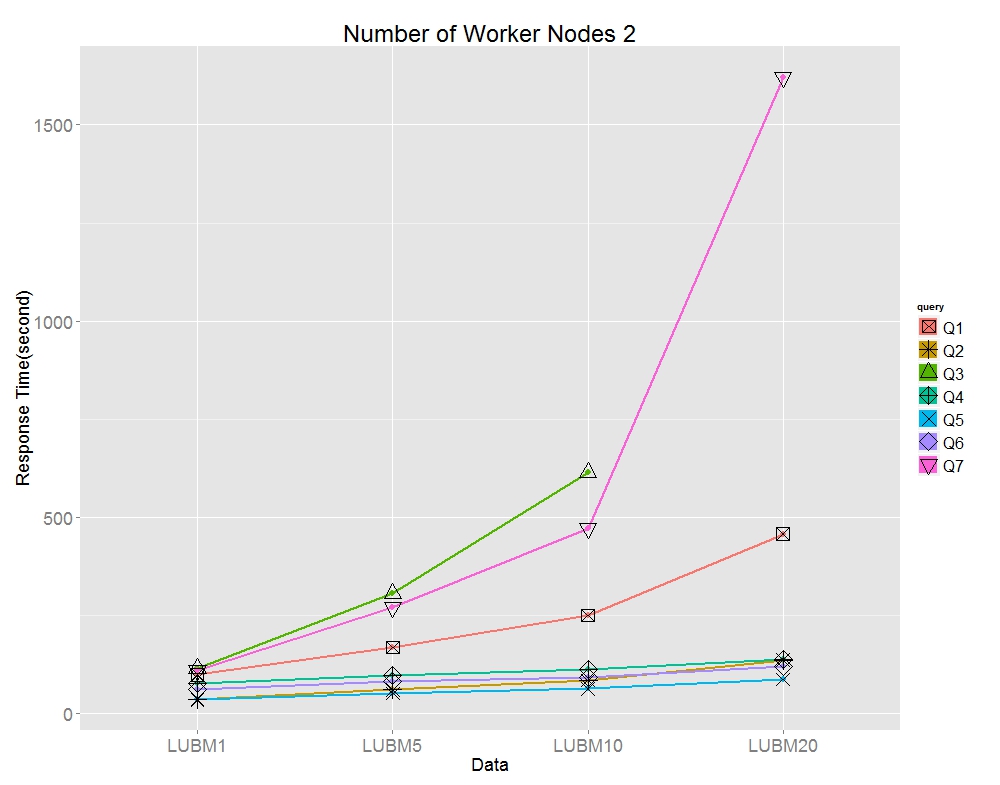}
		\caption{}
		\label{fig:Node2}
	\end{subfigure}
	
	~ 
	\begin{subfigure}[b]{0.45\textwidth}
		\includegraphics[width=\textwidth]{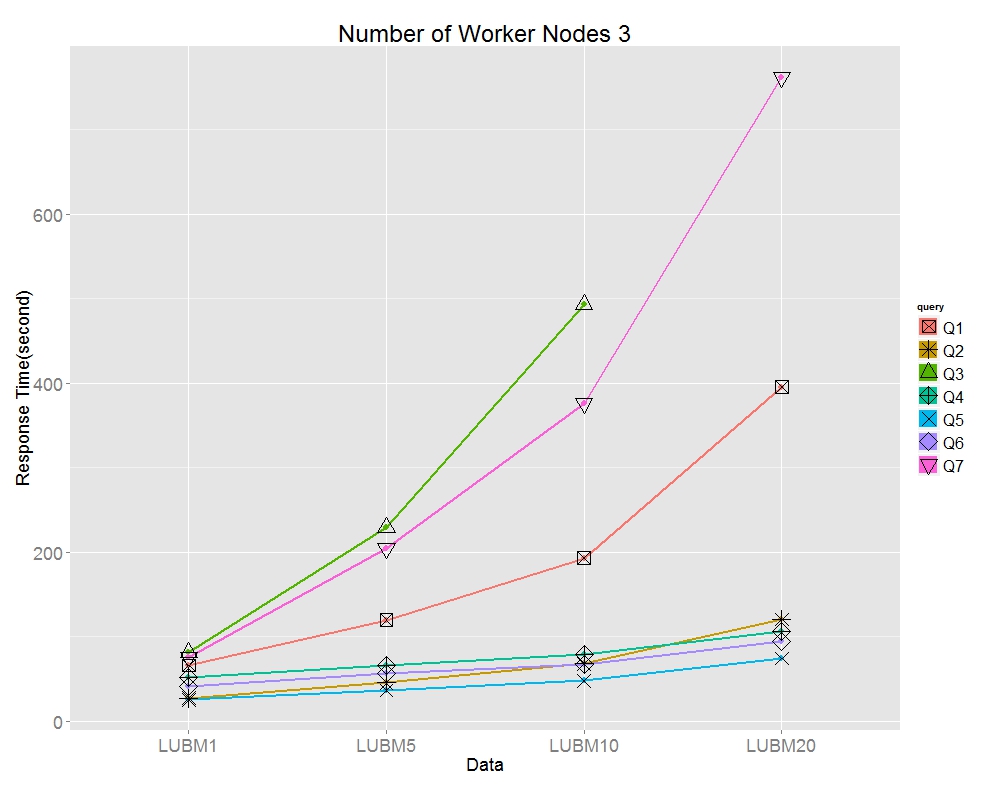}
		\caption{}
		\label{fig:Node3}
	\end{subfigure}
	\qquad	
		\begin{subfigure}[b]{0.45\textwidth}
			\includegraphics[width=\textwidth]{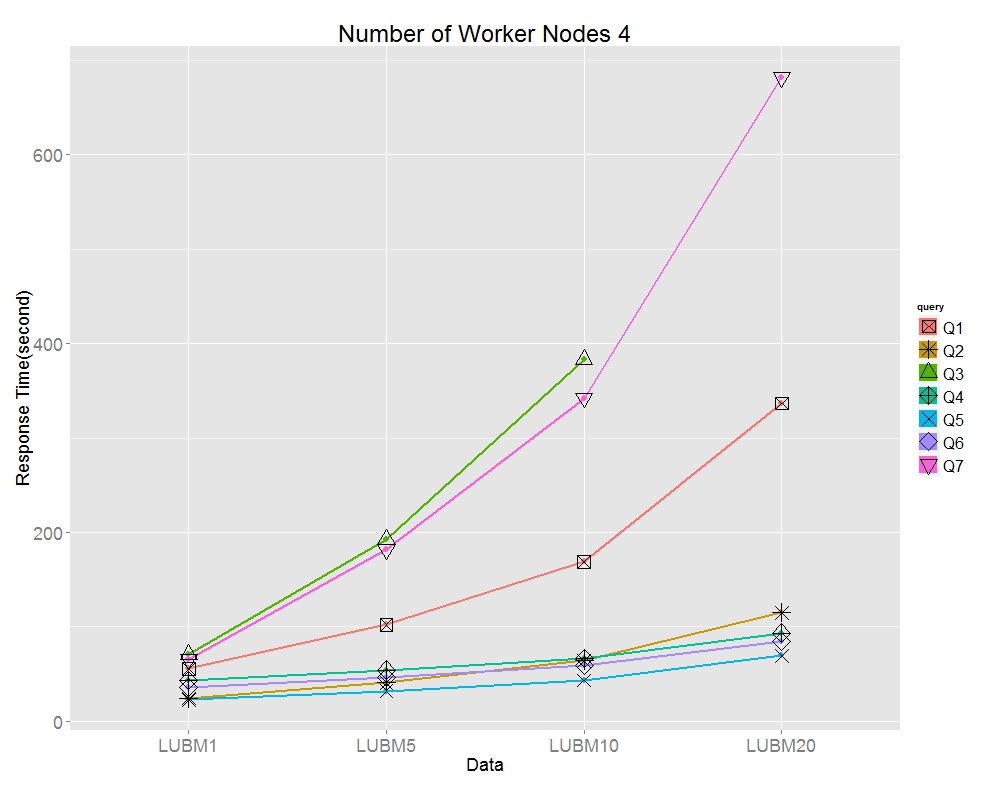}
			\caption{}
			\label{fig:Node4}
		\end{subfigure}

	\begin{subfigure}[b]{0.45\textwidth}
	\includegraphics[width=\textwidth]{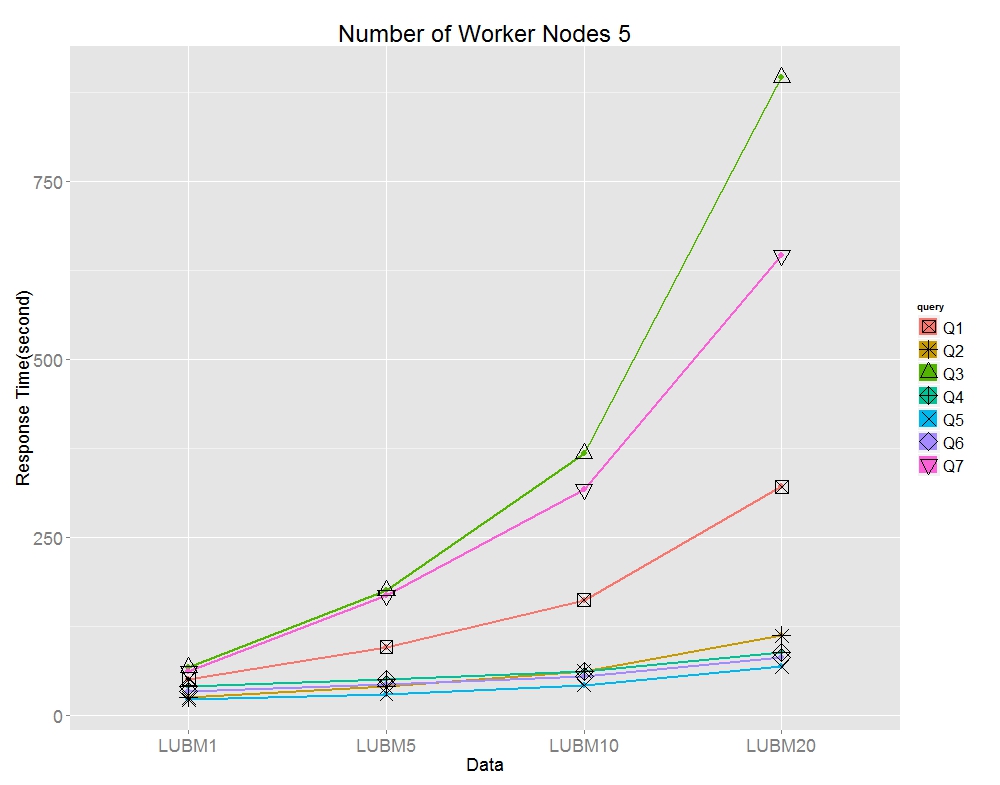}
					\caption{}
					\label{fig:Node5}
				\end{subfigure}
	\caption{Query response time for different data sizes}\label{fig:scalability_nodes}
\end{figure*}

For the less complex, star-join queries (Q2,Q4,Q5,Q6), increasing the data size as the number of nodes is kept constant increases the response time. However, this growth is not considerable in comparison with the more complex, chain queries, as shown in Figure \ref {fig:scalability_nodes}. The first reason is that in our algorithm the number of candidate vertices decreases in each step of evaluating the star queries. This means the sendMsg function executes on fewer vertices after each iteration. As well, the cost of updating vertices after each step, to combine the new and old properties of vertices, is lower. This is because current M\_T tables in these vertices will be empty and the old values are easily replaced with new values. Also, these queries have some constant entities. Constant entities help our algorithm since they limit the number of the vertices with \textit{true} endFlags. This, in turn, leads to fewer RDDs in the final Spark join step.

For more complex queries (Q1, Q3, and Q7), as shown in Figure \ref {fig:scalability_nodes}, the response time increases rapidly as the data size increases. These are chain queries and the selectivity of their triple patterns individually is low. This leads to generate a large number of partial results, which are represented by rows in {M\_T} tables of vertices. Also, these queries have a \textit{cyclic pattern}. So, combining these large {M\_T} tables, resulted from different paths, becomes more costly in vertices that are mapped to a \textit{join-point}. We define a join-point as a vertex, in the query graph, in which different paths meet, e.g. \textit{?z} in Q7. 

However, as shown in Figures \ref{fig:scalability} (a-d), response time of complex queries decreases dramatically as the number of nodes increases. This shows that the system scales well to handle bigger intermediate partial results and the costly vertex updates in these queries.

Furthermore, the results show that currently the system response time is in order of minutes. This happens for some reasons. First, Spark needs more tuning to deal with the proposed algorithm efficiently. Second, the written code suffers from using inefficient data structures of Scala, e.g. \textit{hashmaps}, which consume a lot of memory. In addition, there are some programming patterns in Spark for reducing unnecessary shuffles that need to be considered to improve the system performance.

\section{Practical Experiences}

In this project we found that Spark is difficult to tune since there are quite a number of parameters that need to be dealt with. Also, default values of parameters are not helpful. In fact, Spark's performance is extremely dependent on appropriate parameter values. Among others, it is important to set proper values for \textit{spark.akka.frameSize}, \textit{spark.default.parallelism}, and \textit{spark.executor.memory}. 

The other consideration in programming for Spark is to persist/unpersist RDDs appropriately, i.e., it is important to retain the RDDs in memory only as long as necessary. Otherwise, we may encounter \textit{out of memory} errors. In this work, considering that graph vertices and edges are RDDs, adding new properties to vertices in each iteration generates a new graph. So, we need to unpersist the previous graph and its related RDDs to reduce the memory usage.

Another point is using \textit{shared} variables in Spark. Spark has two types of shared variables: \textit{Broadcast variables} and \textit{Accumulators}. A broadcast variable is used to ship read-only variables to all workers. On the other hand, accumulators support an associative add method to accumulate values in parallel. We use accumulators to build the candidate vertices list in each iteration. Without accumulators we had to collect the current properties of all vertices in \textit{the driver program} which is very expensive operation.

\section{Related Work}
Among all the systems which are designed for dealing with RDF data, we focus on the graph-oriented systems. gStore \cite{zou_gstore:_2013} is a graph-oriented RDF store. This system uses a graph storage model for storing RDF data. gStore transforms RDF data into a so-called \textit{data signature graph}. It implements an efficient subgraph matching mechanism using an \textit{S-tree index} to evaluate SPARQL queries. Also, the signature graph representation enables gStore to answer wildcard queries. 

While gStore is a specialized graph-oriented system for answering SPARQL queries, there are other systems based on some general-purpose graph processing systems. \textit{Trinity.RDF} is a distributed, memory-based engine for RDF data, built on top of \textit{Trinity} graph processing system. It stores data in a native graph form for achieving better performance. This prevents multiple relational-style join to evaluate SPARQL queries. As well, this allows Trininty.RDF to be used for other forms of graph analysis such as reachability, which cannot be expressed by SPARQL. Moreover, Trinity.RDF uses a graph exploration algorithm instead of joins for processing SPARQL queries over its native graph representation. This algorithm treats a query as a sequence of triple patterns instead of evaluating each triple independently as we see in traditional systems. 

Another recent graph-based approach to answer SPARQL queries is \cite{Goodman:2014:UVP:2688283.2688287}. This system is based on GraphLab and employs a vertex-centric subgraph matching algorithm. This work uses the GAS mechanism for implementing the algorithm. However, it has some shortcomings in answering star queries and also it cannot evaluate queries with variables in the position of predicates.

Using Spark for answering SPARQL queries has been investigated briefly in a short paper \cite{DBLP_ChenCZZ14}. The paper introduces \textit{SparkRDF} as a graph processing system for evaluating only SPARQL queries. However, the paper does not provide implementation details of the system.

\section{Conclusion and Future Work}

This work presents a system to evaluate SPARQL BGP queries using GraphX. We introduce a parallel subgraph matching algorithm that is compatible with the GraphX and Spark programming model. GraphX's ability to switch between graph and collection views of the same physical data, Spark's ability in persisting RDDs in memory, and the provided data flow operators over RDDs are crucial in this work. 

Our algorithm represents RDF data as a graph using RDDs in GraphX. By iteratively evaluating BGP triples, the chain and star patterns of a SPARQL query in the RDF graph are tracked and some partial results of subgraph matching are created. The iterative nature of our algorithm relies on Spark's ability to cache intermediate states of graphs, such as the partial results of subgraph matching, in memory. Finally, we switch into a collection view of the partial results. Then, we use Spark as a data-parallel framework to join the partial results for generating final answers. Our experiments, conducted on a small cluster, show the scalability of our proposed algorithm and its promise to handle larger amounts of RDF data by horizontal scaling.

The system still has a preliminary implementation and many enhancements can be projected for improving its performance. Spark is a new framework and many optimizations in our system can be expected by using the new implementation patterns proposed in Spark. As another improvement, we need to use integer IDs instead of string labels for vertices and edges. This prevents from excessive memory usage and improves the join speed. It is also important to apply a query optimization mechanism for generating optimal evaluation orders of BGP triples, as their order affects our system's performance. Furthermore, we first load data into Scala in-memory objects to be loaded into RDDs later. However, this naive approach will be an issue for large graphs that do not fit in memory. As an improvement, we need to implement a mechanism for directly loading graph data from disk into RDDs.

To improve our experiments, it is important to use larger data sets and clusters. Nevertheless, using more sophisticated cluster managers, such as Yarn, and a distributed file system such as Hadoop, would be beneficial. Comparing our system performance to other systems is also one of our goals for extending our work. We currently use GraphX default partitioning strategy. Another experiment is investigating the effect of choosing different partitioning strategies on the system performance.



\bibliographystyle{acm}
\bibliography{refernces}

\newpage

\appendix
\section{GraphX Operators}
\label{App:AppendixA}

Listing ~\ref{lst:operators} presents some important GraphX operators.

\begin{lstlisting}[caption={Summary List of Operators},label={lst:operators}]

class Graph[VD, ED] {
 // Information about the Graph 
  val numEdges: Long
  val numVertices: Long
 // Views of the graph as collections
  val vertices: VertexRDD[VD]
  val edges: EdgeRDD[ED]
  val triplets: RDD[EdgeTriplet[VD, ED]]
 // Functions for caching graphs
  def persist(newLevel: StorageLevel = StorageLevel.MEMORY_ONLY): Graph[VD, ED]
 // Transform vertex and edge attributes
  def mapVertices[VD2](map: (VertexID, VD) => VD2): Graph[VD2, ED]
  def mapEdges[ED2](map: Edge[ED] => ED2): Graph[VD, ED2]
  def mapTriplets[ED2](map: EdgeTriplet[VD, ED] => ED2): Graph[VD, ED2]
 // Modify the graph structure
  def reverse: Graph[VD, ED]
 // Join RDDs with the graph
  def joinVertices[U](table: RDD[(VertexID, U)])(mapFunc: (VertexID, VD, U) => VD): Graph[VD, ED]
 // Aggregate information about adjacent triplets
  def aggregateMessages[Msg: ClassTag]( sendMsg: EdgeContext[VD, ED, Msg] => Unit, mergeMsg: (Msg, Msg) => Msg,tripletFields: TripletFields = TripletFields.All) : VertexRDD[A] }

\end{lstlisting}

\section{LUBM Queries}

The list of LUBM queries which are used in this work.

\label{App:AppendixB}

\textbf{Q1}: SELECT ?x ?y ?z WHERE { ?z ub:subOrganizationOf ?y . ?y rdf:type ub:University . ?z rdf:type ub:Department
. ?x ub:memberOf ?z . ?x rdf:type ub:GraduateStudent . ?x
ub:undergraduateDegreeFrom ?y . }

\textbf{Q2}: SELECT ?x WHERE { ?x rdf:type ub:Course . ?x ub:name
?y . }

\textbf{Q3}: SELECT ?x ?y ?z WHERE { ?x rdf:type ub:UndergraduateStudent.
?y rdf:type ub:University . ?z rdf:type ub:Department .
?x ub:memberOf ?z . ?z ub:subOrganizationOf ?y . ?x ub:undergraduateDegreeFrom
?y . }

\textbf{Q4}: SELECT ?x WHERE { ?x ub:worksFor <http://www.-
Department0.University0.edu> . ?x rdf:type ub:FullProfessor .
?x ub:name ?y1 . ?x ub:emailAddress ?y2 . ?x ub:telephone ?y3
. }

\textbf{Q5}: SELECT ?x WHERE { ?x ub:subOrganizationOf <http-
://www.Department0.University0.edu> . ?x rdf:type ub:ResearchGroup}

\textbf{Q6}: SELECT ?x ?y WHERE { ?y ub:subOrganizationOf <http-
://www.University0.edu> . ?y rdf:type ub:Department . ?x
ub:worksFor ?y . ?x rdf:type ub:FullProfessor . }

\textbf{Q7}: SELECT ?x ?y ?z WHERE { ?y ub:teacherOf ?z . ?y
rdf:type ub:FullProfessor . ?z rdf:type ub:Course . ?x ub:advisor
?y . ?x rdf:type ub:UndergraduateStudent . ?x ub:takesCourse
?z}
\end{document}